\documentclass[prd,aps,eqsecnum,showpacs]{revtex4}

\usepackage{amsmath,amssymb}
\usepackage{bm}
\usepackage[dvips]{graphicx,color}

\begin{document}

\title{Large non-Gaussianity from multi-brid inflation}

\author{Atsushi {\scshape NARUKO} and Misao {\scshape SASAKI}
}

\affiliation{
Yukawa Institute for Theoretical Physics, Kyoto University, 
Kyoto 606-8502, 
Japan
}

\date{\today}

\begin{abstract}
A model of multi-component hybrid inflation,
dubbed multi-brid inflation, in which various observable quantities 
including the non-Gaussianity parameter $f_{NL}$ can be analytically 
calculated was proposed recently.
In particular, for a two-brid inflation model with an exponential
potential and the condition that the end of inflation is
an ellipse in the field space, it was found that,
while keeping the other observational quantities within
the range consistent with observations, large 
non-Gaussianity is possible for certain inflationary trajectories,
provided that the ratio of the two masses is large.
One might question whether the resulting large non-Gaussianity is 
specific to this particular form of the potential and
the condition for the end of inflation.
In this paper, we consider a model of multi-brid inflation
with a potential given by an exponential function 
of terms quadratic in the scalar field components. 
We also consider a more general class of ellipses for 
the end of inflation than those studied previously.
Then, focusing on the case of two-brid inflation,
we find that large non-Gaussianity is possible in the present
model even for the equal-mass case. Then by tuning the model 
parameters, we find that there exist models for which both 
the non-Gaussianity and the tensor-to-scalar ratio
are large enough to be detected in the very near future.
\end{abstract}


\maketitle

\section{Introduction}

The primordial non-Gaussianity has been one of the hottest
topics in cosmology in recent years.
The conventional, single-field slow-roll inflation predicts
that the curvature perturbation is Gaussian to an extremely
high accuracy~\cite{Salopek:1990jq}.
 In other words, if any primordial non-Gaussianity
is detected, it strongly indicates that the dynamics
of inflation is not as simple as we thought it to be.

The primordial non-Gaussianity is conveniently represented 
by a parameter denoted by $f_{NL}$~\cite{Komatsu:2001rj}.
Roughly, it is the ratio of the 3-point correlation function
(or the bispectrum) to the square of the 2-point correlation
function (or the square of the spectrum). It is expected
that near-future experiments such as those of PLANCK will be able 
to detect $f_{NL}$ at a level as small as 5~\cite{Babich:2004yc}.

Finding even a small deviation from Gaussianity will have profound
implications on the theory of the early universe.
Consequently, numerous types of inflationary models that
produce detectable non-Gaussianity have been proposed and
studied~\cite{NGmodels1,NGmodels2,Bernardeau:2002jf,
Dvali:2003em,Kofman:2003nx,Lyth:2005qk,Alabidi,BarnabyCline,Chambers:2007se}.
In terms of the nature of non-Gaussianities, most of these models
can be classified into two categories; those with
non-Gaussianities arising intrinsically from
the quantum fluctuations, and those with non-Gaussianities due
to nontrivial classical dynamics on superhorizon scales. A typical
example of the former is the DBI inflation, in which the
slow-roll condition can be fully violated~\cite{Silverstein:2003hf}.
In this case, the equilateral $f_{NL}$ (denoted by $f_{NL}^{\rm equil}$) 
representing the amplitude of the bispectrum of the equilateral configurations,
is found to play an important role~\cite{DBIinflation}.
On the other hand, in the latter case where non-Gaussianities
are produced on superhorizon scales, by causality 
the local $f_{NL}$ (denoted by $f_{NL}^{\rm local}$)
characterizes the level of the non-Gaussianity.
It is defined in terms of the coefficient in front of 
the second order curvature perturbation~\cite{Komatsu:2001rj},
\begin{eqnarray}
\Phi=\Phi_L+f_{NL}^{\rm local}\Phi_L^2\,,
\label{fNLlocaldef}
\end{eqnarray}
where $\Phi$ is the curvature perturbation on the Newtonian slice
and $\Phi_L$ is its linear, Gaussian part. 

In this paper, we focus on the latter case, that is, we consider
models that may produce a large value of $f_{NL}^{\rm local}$,
for example 10--100. More specifically, we consider hybrid inflation with
multiple inflaton fields, dubbed multi-brid inflation~\cite{Sasaki:2008uc}.
The inflaton fields are assumed to follow the slow-roll
equations of motion, and their fluctuations are assumed to
be Gaussian. In this case, the $\delta N$ formalism is most
useful for the evaluation of the curvature perturbation and
non-Gaussianity~\cite{Sasaki:1995aw,Sasaki:1998ug,Wands:2000dp,
Lyth:2004gb,Lyth:2005fi}.

As in the conventional hybrid inflation, the
inflaton fields are coupled to a water-fall field, and inflation
ends when the inflaton fields satisfy a certain condition that
triggers the instability of the water-fall field.
However, unlike the case of a single inflaton field
in which there is essentially no degree of freedom in
the condition for the end of inflation, there is a substantial
increase in the degree of freedom at the end of inflation
in multi-brid inflation and it widens the viable range of
the parameter space considerably and leads to
the possibility of generating large non-Gaussianity.

As a model of multi-brid inflation, an analytically solvable two-brid
inflation model was recently investigated in detail~\cite{Sasaki:2008uc},
where the potential was assumed to be exponential with
the exponent given by a linear combination of the inflaton fields.
In this paper, we consider a two-brid model with again an exponential potential
but with the exponent given by a quadratic function
of the inflaton fields. The potential has point symmetry about the origin
of the field space, in contrast to the case of the linear exponent 
which has no symmetry. Thus by investigating the quadratic case,
we will be able to see if the generation of large non-Gaussianity
is a generic feature of multi-brid inflation or if it is due to the
lack of symmetry that leads to large non-Gaussianity in the linear
exponent case.

In passing, we mention that the possibility of large non-Gaussianity
from loop correction terms in the perturbative expansion
has recently been discussed by Cogollo et al.~\cite{Cogollo:2008bi}.
Although such a case certainly needs further investigation, in this
paper we concentrate on the case in which leading order (tree) terms
dominate over loop correction terms.

This paper is organized as follows. In \S 2,
we describe our model and derive basic formulas to be used
in the proceeding sections.
In \S 3, using the formulas derived in \S 2, we analytically compute
the spectrum of the curvature perturbation ${\cal P}_S(k)$,
its spectral index $n_S$, the tensor-to-scalar ratio $r$,
and the non-Gaussianity parameter $f_{NL}^{\rm local}$.
In \S 4, as a couple of special cases of our model,
we analyze in detail the equal-mass case as well as the case of
large mass ratio. We find that a large $f_{NL}^{\rm local}$ is possible in
both cases. In particular, in the equal-mass case,
by tuning the parameters to some extent, we find that it is possible to 
have both $r$ and $f_{NL}^{\rm local}$ large enough to be detected.
We conclude the paper in \S 5.
Some computational details are described in Appendix~\ref{app:cal}.
For comparison, we also
summarize the result for the case of the linear exponent two-brid
model in Appendix~\ref{app:linear}.
We use the Planck units where $M_{pl}^{-2}=8\pi G=1$.

\section{Two-brid inflation with approximately quadratic potential}
\label{sec:model}

We consider a two-component scalar field whose action is given by
\begin{eqnarray}
S=-\int d^{4}x\sqrt{-g}
\left[\frac{1}{2}g^{\mu \nu}
\sum_{A=1,2}\partial_{\mu}\phi_A\partial_{\nu}\phi_A
+V(\phi)\right] \,,
\end{eqnarray}
where the potential is given by
\begin{eqnarray}
V=V_0(\chi,\phi_1,\phi_2)
\exp\left[\frac{1}{2}\left(m_1^2\phi_1^2+m_2^2\phi_2^2\right)\right]\,,
\end{eqnarray}
with $V_0$ being a function of a water-fall field $\chi$ as well as
of $\phi_1$ and $\phi_2$, but is assumed to be constant in time
during inflation; see Eq.~(\ref{WFpot}) below.
The same model was previously analyzed~\cite{Sasaki:2007ay}.
However, the condition for the end of inflation considered then
was not general enough to allow the possibility of large non-Gaussianity.

The Friedmann and the field equations are
\begin{gather}
3H^{2}=\frac{1}{2}(\dot\phi_1^2+\dot\phi_2^2)+V(\phi)\,, \\
\ddot {\phi}_A+3H\dot{\phi}_A+\frac{\partial V}{\partial\phi_A}=0\,,
\end{gather} 
where $H=\dot{a}/a$ and the dot~$\dot{~}$ denotes a derivative with
 respect to the cosmic proper time; $\dot{~} =d/dt$.
The slow-roll equations of motion are obtained by neglecting the kinetic term
in the Friedmann equation and the second time derivative in the field equations.
Thus the slow-roll equations of motion are
\begin{eqnarray}
3H^2=V\,,\quad
\frac{d\phi_{A}}{dN}
=\frac{1}{V}\frac{\partial V}{\partial \phi_{A}}=m_{A}^{2}\phi_{A}\,,
\label{slowrolleq}
\end{eqnarray}
where the number of $e$-folds counted backwards in time, $dN=-Hdt$,
is used as the time variable for later convenience.
Note that the effective mass squared $M_{A}^{2}$ for each $\phi_{A}$ is given by
\begin{gather}
M_{A}^{2}=\frac{\partial^{2}V}{\partial \phi_{A}^{2}}
=m_{A}^{2}(1+m_{A}^{2}\phi_{A}^{2})V
=3m_{A}^{2}(1+m_{A}^{2}\phi_{A}^{2})H^{2}\,.
\end{gather}
Thus, the slow-roll condition is satisfied if $m_A^2\ll1$ and 
$\phi_A^2$ are not too much larger than unity. Incidentally, 
since $m_A^2\phi_A^2\ll1$ under this assumption, the difference between the
present potential and a pure quadratic potential,
\begin{eqnarray}
V=V_0+\frac{1}{2}(M_1^2\phi_1^2+M_2^2\phi_2^2)\,,
\end{eqnarray}
is almost negligibly small (that is, they are equivalent to
the leading order in the slow-roll approximation).
An analytical solution for this separable potential model
was first discussed by Starobinsky~\cite{Starobinsky:1986fxa}.

Introducing new field variables $q_{A}$ as
\begin{gather}
\ln q_{1}=\ln (q\cos\theta )
=\int\frac{d\phi_{1}}{m_{1}^{2}\phi_{1}}=\frac{\ln\phi_{1}}{m_{1}^{2}}\,,
\quad
\ln q_{2}=\ln (q\sin\theta )=\frac{\ln\phi_{2}}{m_{2}^{2}}\,,
\end{gather}
the slow-roll equations become 
\begin{gather}
\frac{d\ln q}{dN}=1\,,
\quad
\frac{d\theta}{dN}=0\,.
\end{gather}
Hence we immediately obtain
\begin{gather}
N=\ln q -\ln q_{f}
=\frac{1}{2}\ln\left[\phi_{1}^{2/m_{1}^2}+\phi_{2}^{2/m_{2}^{2}}\right]
-\frac{1}{2}\ln\left[\phi_{1,f}^{2/m_{1}^{2}}+\phi_{2,f}^{2/m_{2}^{2}}\right]\,,
\label{efolds}
\end{gather}
where the number of $e$-folds is set to zero at the end of inflation
and $\phi_{A,f}$ is the final value of the inflaton fields.

\begin{center}
\begin{figure}
\includegraphics[width=10cm]{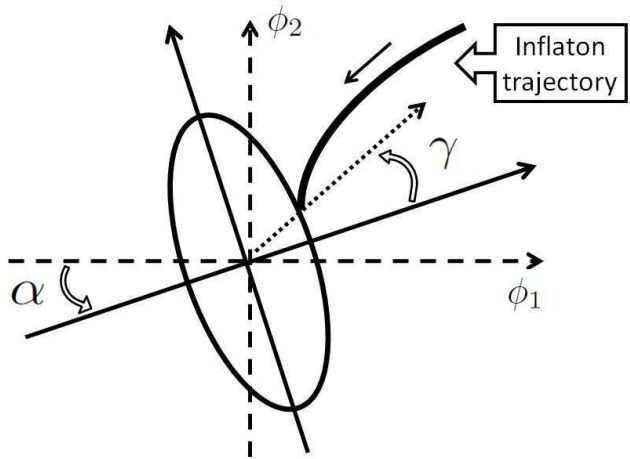}
\caption{Definitions of parameters $\alpha$ and $\gamma$
in field space. The ellipse represents the surface of
the end of inflation. 
}
\label{fig:def}
\end{figure}
\end{center}

We assume that inflation ends at
\begin{align}
\sigma^2=G(\phi_1,\phi_2)\equiv
g_{1}^{2}(\phi_{1}\cos\alpha+\phi_{2}\sin\alpha)^{2}
+g_{2}^{2}(-\phi_{1}\sin\alpha+\phi_{2}\cos\alpha)^{2}\,,
\label{endcondition}
\end{align}
which is realized by the potential $V_0$ given by
\begin{gather}
V_{0}=\frac{1}{2}G(\phi_{1},\phi_{2})\chi^2
+\frac{\lambda }{4}\left( \chi^{2}-\frac{\sigma^{2}}{\lambda}\right)^{2}.
\label{WFpot}
\end{gather}
We parametrize the scalar fields at the end of inflation as
\begin{align}
\frac{\sigma}{g_{1}}\cos\gamma=\phi_{1,f}\cos\alpha+\phi_{2,f}\sin\alpha\,,
\quad
\frac{\sigma}{g_{2}}\sin\gamma=-\phi_{1,f}\sin\alpha+\phi_{2,f}\cos\alpha\,,
\end{align}
namely,
\begin{gather}
\phi_{1,f}
=\frac{\sigma}{g_{1}g_{2}}(g_{2}\cos\alpha \cos\gamma-g_{1}\sin\alpha \sin\gamma),
\quad
\phi_{2,f}
=\frac{\sigma}{g_{1}g_{2}}(g_{2}\sin\alpha \cos\gamma+g_{1}\cos\alpha \sin\gamma).
\label{phifin}
\end{gather}
Figure 1 shows the definitions of the angles $\alpha$ and $\gamma$.
The ellipse describes the surface at the end of inflation,
defined by Eq.~(\ref{endcondition}).
The angle $\alpha$ describes the amount of rotation of
the ellipse relative to the $\phi_1$ and $\phi_2$ axes.
The angle $\gamma$ describes the position of the inflaton
trajectory at the end of inflation.

Since $\theta$ is a constant of motion, we have
\begin{eqnarray}
\ln\left[\frac{q_{1}}{q_{2}}\right]
&=&\frac{1}{m_{1}^{2}}\ln\phi_{1}-\frac{1}{m_{2}^{2}}\ln\phi_{2}
\cr
&=&
\frac{1}{m_{1}^{2}}\ln\frac{\sigma}{g_{1}g_{2}}
(g_{2}\cos\alpha \cos\gamma-g_{1}\sin\alpha \sin\gamma)
-\frac{1}{m_{2}^{2}}\ln\frac{\sigma}{g_{1}g_{2}}
(g_{2}\sin\alpha \cos\gamma-g_{1}\cos\alpha \sin\gamma).
\label{gamma}
\end{eqnarray}
This equation determines the parameter $\gamma$ in terms of 
$\phi_{1}$ and $\phi_{2}$: $\gamma=\gamma(\phi_{1},\phi_{2})$.
Hence, from Eq.~(\ref{phifin}), $\phi_{1,f}$ and $\phi_{2,f}$ become
functions of $\phi_{1}$ and $\phi_{2}$,
\begin{eqnarray}
\phi_{1,f}=\phi_{1,f}(\phi_1,\phi_2)\,,
\quad
\phi_{2,f}=\phi_{2,f}(\phi_1,\phi_2)\,.
\label{phifinf}
\end{eqnarray}
With this understanding, the number of $e$-folds given by Eq.~(\ref{efolds})
becomes a function of $(\phi_{1},\phi_{2})$.
It is then straightforward to obtain $\delta N$ to full nonlinear order. 
It can be straightforwardly calculated as
\begin{gather}
\delta N
=N(\phi_{1}+\delta\phi_{1},\phi_{2}+\delta\phi_{2})-N(\phi_{1},\phi_{2})\,.
\label{deltaN}
\end{gather}

Before closing this section, let us make a small comment.
As mentioned in~\cite{Sasaki:2008uc}, the above formula for $\delta N$
neglects the fact that the surface at the end of inflation,
determined by Eq.~(\ref{endcondition}), is not an equipotential surface.
This will give rise to an additional correction to the final $\delta N$.
Nevertheless, it turns out that the correction is small and can
be neglected, as discussed in~\cite{Sasaki:2008uc}.

\section{Curvature perturbation and non-Gaussianity}
\label{sec:perturb}

In this section, we compute the curvature perturbation of our model
explicitly, and evaluate the curvature perturbation spectrum
${\cal P}_{S}$, the spectral index $n_{S}$, the tensor-to-scalar ratio $r$,
and the non-Gaussianity parameter $f_{NL}^{\rm local}$.

We expand the $\delta N$ formula Eq.~(\ref{deltaN}) to the second order in 
$\delta\phi$ for $N(\phi_1,\phi_2)$ given in Eq.~(\ref{efolds}).
Note that $\delta\gamma$ must be expressed in terms of $\delta\phi$ 
with second-order accuracy.
Details are deferred to Appendix~\ref{app:cal}.
The result is
\begin{eqnarray}
\delta N
&=&\frac{\displaystyle
-\frac{W}{Z}\frac{\delta\phi_{1}}{\phi_{1}}
+\frac{Y}{X}\frac{\displaystyle\delta\phi_{2}}{\phi_{2}}}
{\displaystyle m_{2}^{2}\frac{Y}{X}-m_{1}^{2}\frac{W}{Z}}
\cr
\cr
&&
+\frac{1}{2}\frac{\displaystyle
 \frac{W}{Z}\left(\frac{\delta\phi_{1}}{\phi_{1}}\right)^{2}
-\frac{Y}{X}\left(\frac{\delta\phi_{2}}{\phi_{2}}\right)^{2}}
{\displaystyle m_{2}^{2}\frac{Y}{X}-m_{1}^{2}\frac{W}{Z}}
-\frac{1}{2}\frac{\displaystyle
 \left( 1-\frac{Y}{X}\frac{W}{Z}\right)
\left(\displaystyle  \frac{W}{Z}-\frac{Y}{X}\right)
\left(\frac{\displaystyle m_{2}^{2}}{\phi_{1}}\delta \phi_{1}
-\frac{\displaystyle m_{1}^{2}}{\phi_{2}}\delta \phi_{2}\right)^{2}}
{\left(\displaystyle m_{2}^{2}\frac{Y}{X}-m_{1}^{2}\frac{W}{Z}\right)^{3}}
+\cdots\,,
\label{dN2nd}
\end{eqnarray}
where we have introduced the quantities and defined as
\begin{gather}
g=\sqrt{g_1^2+g_2^2}\,,
\quad
g_1=g\cos\beta\,, 
\quad
g_2=g\sin\beta\,,
\notag\\
X=\frac{1}{g}(g_{2}\cos\alpha \cos\gamma -g_{1}\sin\alpha \sin\gamma)
\propto\phi_{1,f}\,,
\quad
Y=\frac{1}{g}(g_{2}\cos\alpha \sin\gamma +g_{1}\sin\alpha \cos\gamma)
=-\frac{\partial}{\partial\gamma}X\,,
\notag\\
Z=\frac{1}{g}(g_{2}\sin\alpha \cos\gamma +g_{1}\cos\alpha \sin\gamma)
\propto\phi_{2,f}\,,
\quad
W=\frac{1}{g}(g_{2}\sin\alpha \sin\gamma -g_{1}\cos\alpha \cos\gamma)
=-\frac{\partial}{\partial\gamma}Z\,.
\notag
\end{gather}
We note that $\tan\beta=g_2/g_1$. For example, in the case
of Fig.~1, $\tan\beta$ is the ratio of the semiminor axis to
the semimajor axis (hence $\beta<\pi/4$).

We assume that the scalar field fluctuations $\delta\phi_{1}$ and 
$\delta\phi_{2}$ are Gaussian with the dispersion,
\begin{eqnarray}
\left\langle\delta\phi_{A}\delta\phi_{B}\right\rangle_{k}
=\left( \frac{H}{2\pi}\right)^{2}_{t_{k}}\delta_{AB}\,,
\label{Gaussian}
\end{eqnarray}
where $t_{k}$ is the horizon-crossing time of the comoving wave number $k$,
where $k=Ha$. Then the curvature perturbation spectrum is given by
\begin{eqnarray}
{\cal P}_{S}(k)\equiv \frac{4\pi k^{3}}{(2\pi)^{3}}P_{\cal{R}}(k)
=\frac{g^{2}}{\sigma^{2}}
\frac{\sin^{2}\beta\cos^{2}\beta}{(m_{2}^{2}\,YZ-m_{1}^{2}\,XW)^{2}}
\left(W^{2}e^{-2m_{1}^{2}N_k}+Y^{2}e^{-2m_{2}^{2}N_k}\right)
\left(\frac{H}{2\pi}\right)^{2}_{t_{k}}\,,
\label{PS}
\end{eqnarray}
where $N_k$ is the number of $e$-folds at the horizon crossing, $N_k=N(t_k)$.
Using the fact that $3H^{2}=V$, the spectral index is found to be 
\begin{eqnarray}
n_{S}-1
=2\frac{m_{1}^{2}e^{-2m_{1}^{2}N_k}W^{2}+m_{2}^{2}e^{-2m_{2}^{2}N_k}Y^{2}}
{e^{-2m_{1}^{2}N_k}W^{2}+e^{-2m_{2}^{2}N_k}Y^{2}}
-\frac{\sigma^{2}}{g^{2}}
\frac{m_{1}^{4}e^{2m_{1}^{2}N_k}X^{2}+m_{2}^{4}e^{2m_{2}^{2}N_k}Z^{2}}
{\sin^{2}\beta\cos^{2}\beta}\,.
\label{nS}
\end{eqnarray}
The tensor-to-scalar ratio is given by
\begin{eqnarray}
r \equiv \frac{{\cal P}_{T}}{{\cal P}_{S}}
=8\frac{\sigma^{2}}{g^{2}}
\frac{(m_{2}^{2}YZ-m_{1}^{2}XW)^{2}}{\sin^{2}\beta\cos^{2}\beta}
\frac{1}{W^{2}e^{-2m_{1}^{2}N_k}+Y^{2}e^{-2m_{2}^{2}N_k}}\,.
\label{TSr}
\end{eqnarray}

Now we evaluate the non-Gaussianity in our model. 
For convenience, we introduce the linear curvature perturbation
${\cal R}_{L}$ and the linear entropy perturbation $S$,
\begin{gather}
{\cal R}_{L}
=\frac{\displaystyle -\frac{XW}{\phi_{1}}\delta \phi_{1}
+\frac{YZ}{\phi_{2}}\delta\phi_{2}}{m_{2}^{2}\,YZ-m_{1}^{2}\,XW}\,,
\quad
S=\frac{\displaystyle \frac{YZ}{\phi_{2}}\delta \phi_{1}
+\frac{XW}{\phi_{1}}\delta\phi_{2}}{m_{2}^{2}\,YZ-m_{1}^{2}\,XW}\,.
\end{gather}
For the Gaussian fluctuations $\delta\phi_{A}$ given by Eq.~(\ref{Gaussian}),
we see that $S$ has the same spectrum as the curvature perturbation
${\cal R}_L$, but is orthogonal to it,
\begin{gather}
\left\langle{\cal R}_{L}\cdot S\right\rangle=0\,.
\end{gather}
In terms of ${\cal R}_{L}$ and $S$, the nonlinear $\delta N$ in 
Eq.~(\ref{dN2nd}) is re-expressed as
\begin{gather}
\delta N
={\cal R}_{L}+\frac{3}{5}f_{NL}^{\rm local}\,{\cal R}_{L}^{2}
+C_{RS}\,{\cal R}_L\,S+C_{SS}\,S^2+\cdots\,,
\end{gather}
where the non-Gaussian parameter $f_{NL}^{\rm local}$ is given by
\begin{eqnarray}
&&f_{NL}^{\rm local}=XZ
\cr
\cr
&&~\times
\frac{5}{6}\frac{\left(\displaystyle \frac{W^{3}}{X}e^{4m_{2}^{2}N_k}
-\frac{Y^{3}}{Z}e^{4m_{1}^{2}N_k}\right)
\left(\displaystyle  m_{2}^{2}\frac{Y}{X}-m_{1}^{2}\frac{W}{Z}\right)^{2}
-\left(XZ-YW\right)\left(\displaystyle \frac{W}{Z}-\frac{Y}{X}\right)
\left(\displaystyle  m_{1}^{2}\frac{Y}{Z}e^{2m_{1}^{2}N_k}
+m_{2}^{2}\frac{W}{X}e^{2m_{2}^{2}N_k}\right)^{2}}
{\left(\displaystyle  m_{2}^{2}\frac{Y}{X}-m_{1}^{2}\frac{W}{Z}\right)
\left( Y^{2}e^{2m_{1}^{2}N_k}+W^{2}e^{2m_{2}^{2}N_k}\right)^{2}}\,.
\cr
&&\label{f_NL}
\end{eqnarray}
This is one of our main results. 

Before closing this section, let us
briefly discuss the non-Gaussianity due to
the other second order coefficients $C_{RS}$ and $C_{SS}$. 
Although the derivation of their expressions is straightforward,
we do not present their explicit expressions here, because
they are as complicated as Eq.~(\ref{f_NL}) and because they are unnecessary
for the purpose of the present paper.
We only mention that an inspection of the resulting expressions reveals
that they can never become much larger than $f_{NL}^{\rm local}$.
To be a bit more precise, their values can become large 
for certain ranges of the parameters, but when they
become large, $f_{NL}^{\rm local}$ also becomes large
simultaneously. Hence, as discussed in~\cite{Sasaki:2008uc},
as long as we focus on the bispectrum, we can neglect their contribution.

\section{Case for large non-Gaussianity}
\label{sec:f_NL}

Since Eq.~(\ref{f_NL}) for $f_{NL}^{\rm local}$
is very complicated, it is not easy to study all possible cases in detail.
However, there are some limiting cases in which we have a substantially
simplified expression for $f_{NL}^{\rm local}$ but which are yet 
sufficiently of interest.

One case of interest is when the two masses are equal, $m_1=m_2$.
In this case, the potential during inflation is \textit{O}(2) symmetric.
This symmetry is broken at the end of inflation because of  
condition~(\ref{endcondition}), unless $g_1=g_2$. This model was 
discussed by Alabidi and Lyth~\cite{Alabidi} as a new mechanism of
generating curvature perturbations. 
Another case of interest is when the ratio of
the mass parameters are large, for example, $m_{1}\gg m_{2}$.
In this large mass ratio limit, an inspection of the Eq.~(\ref{f_NL})
suggests that a large value of $f_{NL}^{\rm local}$
may be possible if the parameter $W$ is very small.
In this section, we investigate these two cases in detail.

\subsection{Equal mass}
\label{ss:equalmass}

First let us consider the equal mass case, $m_1^2=m_2^2\equiv m^2$.
This means that there is \textit{O}(2) symmetry during inflation,
and the symmetry is spontaneously broken at the end of
inflation~\cite{Alabidi}.

In the equal-mass case, the formulas derived in the
previous section simplify considerably to
\begin{eqnarray}
{\cal{P}}_S
&=&\left(\frac{g}{\sigma m^{2}e^{m^{2}N_k}}\right)^{2}
\frac{1+\cos 2\beta \cos 2\gamma}{2}
\left(\frac{H}{2\pi}\right)^{2}_{t_{k}}
=\frac{8}{r}\left(\frac{H}{2\pi}\right)^{2}_{t_{k}}\,,
\\
n_{S}-1
&=&2m^{2}-2\left(\frac{\sigma m^{2}e^{m^{2}N_k}}{g}\right)^{2}
\frac{1-\cos 2\beta \cos 2\gamma}{\sin^22\beta}
=2m^{2}-\frac{r}{8}\,\frac{1-\cos^22\beta \cos^22\gamma}{\sin^22\beta}\,,
\\
r&=&
8\left(\frac{\sigma m^{2}e^{m^{2}N_k}}{g}\right)^{2}
\frac{2}{1+\cos 2\beta \cos 2\gamma},\,
\\
f_{NL}^{\rm local}
&=&\frac{5m^{2}}{6}
\left\{\left(\frac{\cos{2\beta}\sin{2\gamma}}{1+\cos{2\beta}\cos{2\gamma}}
\right)^{2}-1\right\}\,.
\label{fNLeqmass}
\end{eqnarray}
Note that the $\alpha$-dependence has disappeared because of
the symmetry.

As is clear from Eq.~(\ref{fNLeqmass}) for $f_{NL}^{\rm local}$,
in order to obtain large non-Gaussianity,
it is necessary for the factor in the curly brackets to become large,
that is, $\cos2\beta\sin2\gamma/(1+\cos{2\beta}\cos{2\gamma})\gg1$.
This is possible either in the limit $(\beta,\gamma)\to(0,\pi/2)$ 
or $(\beta,\gamma)\to(\pi/2,0)$. Since these two limits are equivalent,
let us take the limit $(\beta,\gamma)\to(0,\pi/2)$.
This corresponds to the situation in which
the ellipse is highly elongated and the inflaton trajectory hits
the ellipse close to one of the tips of the majoraxis.
Then, setting $\pi/2-\gamma=\delta$, we obtain
\begin{eqnarray}
{\cal{P}}_S
&=&\frac{8}{r}\left(\frac{H}{2\pi}\right)^{2}_{t_{k}}\,,
\\
n_{S}-1
&=&2m^{2}-\frac{r}{8}\,\left(1+\frac{\delta^2}{\beta^2}\right)\,,
\label{nseqm0}\\
r&=&8\left(\frac{\sigma m^{2}e^{m^{2}N}}{g}\right)^{2}
\frac{1}{\beta^2}\left(1+\frac{\delta^2}{\beta^2}\right)^{-1}\,,
\label{reqm0}
\\
f_{NL}^{\rm local}
&=&\frac{5m^{2}}{6}\frac{1}{\beta^2}\frac{\delta^2}{\beta^2}
\left(1+\frac{\delta^2}{\beta^2}\right)^{-2}\,.
\label{fNLeqm0}
\end{eqnarray}

To investigate in more detail the theoretical predictions of this model,
let us derive expressions for $r$ and $f_{NL}^{\rm local}$ in terms
of the observational data as much as possible.
We first fix the amplitude of the spectrum ${\cal P}_S$.
The WMAP normalization~\cite{Liddle:2006ev} gives
\begin{eqnarray}
{\cal P}_S=\frac{8}{r}\left(\frac{H}{2\pi}\right)_{t_k}^2
= 2.5\times 10^{-9}\,,
\label{WMAPnorm}
\end{eqnarray}
at the present Hubble horizon scale. Also, the WMAP 5-year
analysis~\cite{Dunkley:2008ie,Komatsu:2008hk} gives
the spectral index,
\begin{eqnarray}
n_S=0.96~\lower1ex\hbox{$\stackrel{\displaystyle+0.014}{-0.015}$}\,.
\label{nSWMAP}
\end{eqnarray}
Below we replace ${\cal P}_S$ and $n_S$ by these observed values.

Noting the fact that
\begin{eqnarray}
3H^2=V\simeq V_0=\frac{\sigma^4}{4\lambda}\,,
\end{eqnarray}
we obtain, from Eq.~(\ref{WMAPnorm}), the equation,
\begin{eqnarray}
\sigma^4=6\pi^2\times {\cal P}_S\,\lambda\,r\,.
\end{eqnarray}
Inserting this into the square of Eq.~(\ref{reqm0}), we obtain
\begin{eqnarray}
r=384\pi^2{\cal P}_S\frac{\lambda}{g^{4}}\,
\frac{m^{8}e^{4m^{2}N_{k}}}{\beta^{4}}
\left(1+\frac{\delta^{2}}{\beta^{2}}\right)^{-2}\,.
\label{rexp}
\end{eqnarray}
Eliminating $r$ from Eq.~(\ref{nseqm0}) by using Eq.~(\ref{rexp})
gives
\begin{eqnarray}
\left(1+\frac{\delta^2}{\beta^2}\right)
=\frac{48\pi^2{\cal P}_S}{(2m^2-(n_S-1))}\,\frac{\lambda}{g^{4}}\,
\frac{m^{8}e^{4m^{2}N_{k}}}{\beta^{4}}\,.
\label{dbexp}
\end{eqnarray}
Plugging this back into Eq.~(\ref{rexp}),
we obtain the expression for $r$,
\begin{eqnarray}
r=\frac{(2m^2-(n_S-1))^2}{6\pi^2{\cal P}_S}
\frac{g^4}{\lambda}\frac{\beta^4}{m^8e^{4m^2N_k}}\,.
\end{eqnarray}
Also, plugging Eq.~(\ref{dbexp}) into Eq.~(\ref{fNLeqm0}),
we obtain
\begin{eqnarray}
f_{NL}^{\rm local}=\frac{5\,m^2}{6\,\beta^2}
\left(\frac{(2m^2-(n_S-1))}{48\pi^2{\cal P}_S}\,\frac{g^{4}}{\lambda}\,
\frac{\beta^{4}}{m^{8}e^{4m^{2}N_{k}}}\right)^2
\left(\frac{48\pi^2{\cal P}_S}{(2m^2-(n_S-1))}\,\frac{\lambda}{g^{4}}\,
\frac{m^{8}e^{4m^{2}N_{k}}}{\beta^{4}}-1\right)\,.
\end{eqnarray}
Assuming $\delta^2\gg\beta^2$, and using the observed values given in
Eqs.~(\ref{WMAPnorm}) and (\ref{nSWMAP}), the above expressions for 
$r$ and $f_{NL}^{\rm local}$ reduce to
\begin{eqnarray}
r&\sim&2.7\times10^7(m^2+0.02)^2\,\frac{g^4}{\lambda}\,
\frac{\beta^4}{m^8e^{4m^2N_k}}\,,
\label{rnum}
\\
\cr
f_{NL}^{\rm local}
&\sim&1.4\times10^6(m^2+0.02)\,\frac{g^4}{\lambda}\,
\frac{\beta^2}{m^6e^{4m^2N_k}}\,.
\label{fNLnum}
\end{eqnarray}
Another useful expression may be obtained by combining the
above two expressions:
\begin{eqnarray}
f_{NL}^{\rm local}\sim52\left(\frac{r}{0.1}\right)
\left(\frac{10^{-4}}{\beta^2}\right)\frac{m^2}{m^2+0.02}\,.
\end{eqnarray}
This tells us that for $m^2\gtrsim0.02$, in the very near future, 
both $r$ and $f_{NL}^{\rm local}$ may be large enough to be
detected upon tuning the model parameters to some extent.

\begin{center}
\begin{figure}
\includegraphics[width=10cm]{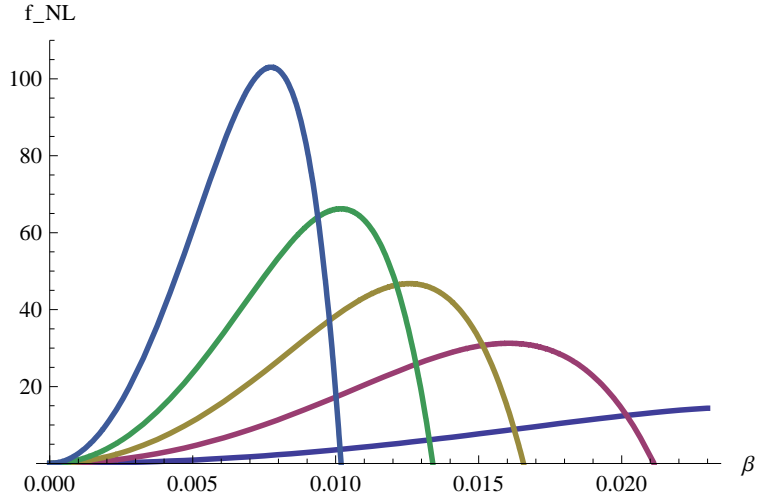}
\caption{Non-Gaussian parameter $f_{NL}^{\rm local}$ as a 
function of $\beta$ for several different values of $m^2$.
The coupling constant parameters are set to $\lambda/g^4=1$.
The spectral index is set to $n_S=0.96$. 
The curves are, from the one with the 
highest peak to that with the lowest peak, for $m^2=1/20$, $1/23$, 
$1/25$, $1/27$, and $1/30$.
}\label{fig:fNL}
\end{figure}
\end{center}

\begin{center}
\begin{figure}
\includegraphics[width=10cm]{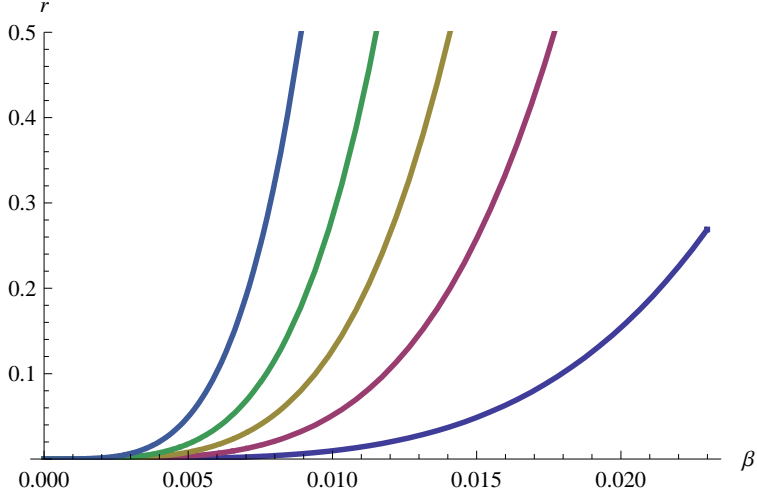}
\caption{Tensor-to-scalar ratio $r$ as a functions of $\beta$
for several different values of $m^2$. The other parameters are the
same as in Fig.~\ref{fig:fNL}. The curves are, from the left to
the right, for $m^2=1/20$, $1/23$, $1/25$, $1/27$, and $1/30$.}
\label{fig:r}
\end{figure}
\end{center}

In Figs.~\ref{fig:fNL} and \ref{fig:r}, we show $f_{NL}^{\rm local}$ 
and $r$, respectively, as functions of $\beta$ for several different 
values of $m^2$. The coupling constants are set to $\lambda/g^4=1$.
The spectral index is set to $n_S=0.96$, but we find the dependence
of it is weak in the range $0.94\lesssim n_S\lesssim0.98$. 
In Fig.~\ref{fig:fNL}, each curve up
to its peak is well approximated by Eq.~(\ref{fNLnum}).
In both figures, if we vary $\lambda/g^4$, the curves will 
scale inversely proportional to $\lambda/g^4$.
As we can see, although the values of $f_{NL}^{\rm local}$ 
and $r$ are relatively sensitive to the values of $m^2$ and
$\beta$, there indeed exist models with large $f_{NL}^{\rm local}$ 
and $r$ simultaneously.


\subsection{Large mass ratio}
\label{ss:largeratio}

Here, we consider the case of large mass ratio.
Let us tentatively assume that $m_1^2\gg m_2^2$.
Then an inspection of Eq.~(\ref{f_NL}) for $f_{NL}^{\rm local}$
suggests that a large $f_{NL}^{\rm local}$ is possible if $W\ll1$.
Hence, let we set $W=0$ for simplicity and investigate this case in detail.
We note that the only assumption we adopt is the condition $W=0$; 
we do not assume a large mass ratio in the following analysis.
Namely, the formulas derived below are valid for any mass ratio
unless otherwise stated.

The condition $W=0$ implies the following relation between the model parameters:
\begin{gather}
\frac{g_{2}}{g_{1}}\tan\alpha\tan\gamma
=\tan\alpha\tan\beta\tan\gamma=1\,.
\label{Weqzero}
\end{gather}
If $\tan\beta=1$, this means that the inflaton trajectory arrives 
at the ellipse along the $\phi_2$ axis.
In this case, Eqs.~(\ref{PS}), (\ref{nS}), (\ref{TSr}), and (\ref{f_NL})
respectively reduce to, 
\begin{eqnarray}
{\cal P}_S
&=&\left(\frac{g}{\sigma\, m_{2}^{2}\,e^{m_{2}^{2}N_k}}\right)^{2}
\frac{1+\cos 2\beta \cos 2\gamma}{2}
\left(\frac{H}{2\pi}\right)^{2}_{t_{k}}
\cr
\cr
&=&\frac{8}{r}\left(\frac{H}{2\pi}\right)^{2}_{t_{k}}\,,
\label{PSW}\\
\cr
n_{S}-1
&=&2\,m_{2}^{2}
-2\frac{\sigma^{2}}{g^{2}}
\frac{m_{1}^{4}e^{2m_{1}^{2}N_k}\cot^22\beta\sin^22\gamma
+m_{2}^{4}e^{2m_{2}^{2}N_k}}{1+\cos 2\beta \cos 2\gamma}
\cr
\cr
&=&2\,m_{2}^{2}
-\frac{r}{8}\left(\frac{m_{1}^{4}}{m_{2}^{4}}
e^{2(m_{1}^{2}-m_{2}^{2})N_k}\cot^2 2\beta\sin^22\gamma+1\right)\,,
\label{nSW}\\
\cr
r&=&8\left(\frac{\sigma\,m_{2}^{2}\,e^{m_{2}^{2}N_k}}{g}\right)^{2}
\frac{2}{1+\cos 2\beta \cos 2\gamma}\,,
\label{rW}\\
\cr
f_{NL}^{\rm local}
&=&\frac{5}{6}\left\{\frac{m_{1}^{4}}{m_{2}^{2}}\left(
\frac{\cos 2\beta \sin 2\gamma}{1+\cos 2\beta \cos 2\gamma}\right)^{2}
-m_{2}^{2}\right\}
\cr
\cr
&=&\frac{5\,m_2^2}{6}\left\{\frac{m_{1}^{4}}{m_{2}^{4}}\left(
\frac{\cos 2\beta \sin 2\gamma}{1+\cos 2\beta \cos 2\gamma}\right)^{2}
-1\right\}
\,.
\label{fNLW}
\end{eqnarray}
We note that the equal-mass limit discussed in the previous subsection
can be obtained by setting $m_1^2=m_2^2=m^2$ in the above equations, 
because the condition $W=0$ becomes irrelevant in the equal mass limit.

Eq.~(\ref{fNLW}) implies that we may have large non-Gaussianity
if $m_1^2\gg m_2^2$ and/or 
$\cos2\beta\sin2\gamma/(1+\cos2\beta\cos2\gamma)\gg1$.
We also note that in both cases the value of 
$f_{NL}^{\rm local}$ will be positive.
This result is the same as the equal mass case and similar to the case
of the linear exponent potential model discussed in~\cite{Sasaki:2008uc}.
We suspect that this positivity property may be generically true for
all models that are capable of producing large {\it local\/}
non-Gaussianity.

First, let us assume that $\cos2\beta\sin2\gamma/(1+\cos2\beta\cos2\gamma)$
is of the order of unity. Recall that we have $m_1^2\ll1$ and $m_2^2\ll1$
from the slow-roll condition.
Then in order to obtain a large $f_{NL}^{\rm local}$, say
$f_{NL}^{\rm local}\sim 50$, we need to have an extremely large
mass ratio, $m_1^2/m_2^2\sim 50\,m_1^{-2}\gg1$. 
Then, Eq.~(\ref{nSW}) implies that $r$ must be extremely small, since
we must have $n_S-1\ll1$. Therefore, large non-Gaussianity is possible 
only in models of very low energy inflation.

In order to look for the possibility of both large $r$ and
large $f_{NL}^{\rm local}$, we consider the case of
$\cos2\beta\sin2\gamma/(1+\cos2\beta\cos2\gamma)\gg1$.
As discussed in the previous subsection, this is realized 
either for $(\beta,\gamma)\to(0,\pi/2)$ or
$(\beta,\gamma)\to(\pi/2,0)$. Again, since both limits are equivalent,
we focus on the limit $(\beta,\gamma)\to(0,\pi/2)$.
In this limit, setting $\pi/2-\gamma=\delta$ again, we have
\begin{eqnarray}
{\cal P}_S
&=&\frac{8}{r}\left(\frac{H}{2\pi}\right)^{2}_{t_{k}}
\sim\left(\frac{g}{\sigma\, m_{2}^{2}\,e^{m_{2}^{2}N_k}}\right)^{2}
\frac{1}{\beta^2+\gamma^2}\left(\frac{H}{2\pi}\right)^{2}_{t_{k}}
\,,
\\
\cr
n_{S}-1
&\sim&2\,m_{2}^{2}-\frac{r}{8}\left(\frac{m_{1}^{4}}{m_{2}^{4}}
e^{2(m_{1}^{2}-m_{2}^{2})N_k}\frac{\delta^2}{\beta^2}+1\right)\,,
\label{nSzero}\\
\cr
r&\sim&8\left(\frac{\sigma\,m_{2}^{2}\,e^{m_{2}^{2}N_k}}{g}\right)^{2}
\frac{1}{\beta^2+\delta^2}\,,
\\
\cr
f_{NL}^{\rm local}
&\sim&
\frac{5}{6}m_2^2\,\frac{m_{1}^{4}}{m_{2}^{4}}
\left(\frac{\delta}{\beta^2+\delta^2}\right)^2\,.
\label{fNLzero}
\end{eqnarray}
Now, if we have $\beta\ll\delta\ll1$, we can obtain large 
$f_{NL}^{\rm local}$. However, again, Eq.~(\ref{nSzero}) 
for $n_S-1$ implies $r$ must be extremely small 
if $m_1^2\gg m_2^2$. In other words, having a large mass ratio
does not help in enlarging the parameter region in which
both $r$ and $f_{NL}^{\rm local}$ are large.

\section{conclusion}
\label{sec:conclusion}

We analytically investigated the curvature perturbation and its
non-Gaussianity in a model of multi-field hybrid inflation, 
dubbed {\it multi-brid\/} inflation. The model we considered
is a two-field hybrid inflation (two-brid inflation) model with 
the potential mimicking conventional quadratic potentials.
The new ingredient of the model is the generalization of the condition
for the end of inflation. We considered a very general coupling
of the two inflaton fields to a water-fall field. 

Then, using the $\delta N$ formula, we derived an
analytical expression for the curvature perturbation.
Based on this expression, we obtained the curvature perturbation
spectrum ${\cal P}_S$, the spectral index $n_S$, the tensor-to-scalar
ratio $r$, and the non-Gaussian parameter $f_{NL}^{\rm local}$.
We found that a large positive $f_{NL}^{\rm local}$ is possible
in this model.
Then, at least for a certain limited range of the parameters,
we explicitly showed that it is possible to have large non-Gaussianity
while keeping the values of the other quantities
consistent with those of the observation. In particular, we showed that
when the two inflaton masses are equal,
the parameters can be tuned so that they lead to a fairly large
tensor-to-scalar ratio, $r\sim 0.1$, as well as a large non-Gaussian
parameter, $f_{NL}^{\rm local}\sim 50$. These values will be 
at a detectable level in the very near future. 
On the other hand, interestingly, we found that having a large mass 
ratio in the present model does not help in producing both 
$r$ and $f_{NL}^{\rm local}$ large enough to be detected.
This is in contrast to the model studied in~\cite{Sasaki:2008uc}.

The standard lore has been that $f_{NL}^{\rm local}$ is too small 
for models with large $r$ or vice versa.
We have shown, in this paper, not be the case, particularly 
in this model of spontaneously symmetry breaking at
the end of inflation. 
This may be the most important conclusion of this work.
At the moment, we have no clear physical explanation 
for this result. We hope we will be able to answer this question
in the near future.

\acknowledgements

The main part of this work was carried out during the international 
molecule-type program, ``Inflationary Cosmology'', under the Yukawa
International Program for Quark-Hadron Sciences.
We would like to thank R. Kallosh and A. Linde, who were the core
participants in the program, for fruitful and illuminating discussions.
This work was also supported in part by JSPS Grants-in-Aid for Scientific 
Research (B) No.~17340075, and (A) No.~18204024, by JSPS 
Grant-in-Aid for Creative Scientific Research No.~19GS0219,
and by Monbukagaku-sho Grant-in-Aid for the global COE program,
``The Next Generation of Physics, Spun from Universality and Emergence''.

\appendix

\section{$\bm{\delta N}$ to Second Order}
\label{app:cal}

Here we evaluate $\delta N$ to the second order in the perturbation.
We assume the field fluctuations $\delta\phi_{1}$ and $\delta\phi_{2}$ are of
linear order.

First, we express the perturbation in the orbital parameter
$\gamma$ in terms of  $\delta\phi_{1}$ and $\delta\phi_{2}$.
Setting $\delta\gamma=\delta_{1}\gamma+\delta_{2}\gamma$, 
where $\delta_{1}\gamma$ and $\delta_{2}\gamma$ are of linear and second orders,
respectively, we take the perturbation of Eq.~(\ref{gamma}) to the second order.
We obtain
\begin{eqnarray}
&&\left(\frac{1}{m_{1}^{2}}\frac{\delta \phi_{1}}{\phi_{1}}
-\frac{1}{2m_{1}^{2}}\frac{\delta\phi_{1}^{2}}{\phi_{1}^{2}}\right)
-\left(\frac{1}{m_{2}^{2}}\frac{\delta \phi_{2}}{\phi_{2}}
-\frac{1}{2m_{2}^{2}}\frac{\delta\phi_{2}^{2}}{\phi_{2}^{2}} \right)
\nonumber\\
&&\quad
=\left(\frac{1}{m_{1}^{2}}\frac{\partial }{\partial \gamma }\ln\phi_{1,f}
-\frac{1}{m_{2}^{2}}\frac{\partial }{\partial \gamma }\ln\phi_{2,f}\right)
(\delta_{1} \gamma +\delta_{2} \gamma)
+\frac{1}{2}\left(\frac{1}{m_{1}^{2}}
\frac{\partial ^{2}}{\partial \gamma ^{2} }\ln\phi_{1,f}
-\frac{1}{m_{2}^{2}} \frac{\partial ^{2}}{\partial \gamma ^{2} }
\ln\phi_{2,f}\right)(\delta_{1} \gamma )^{2}.
\label{expand}
\end{eqnarray}

The linear part of the above equation determines $\delta_{1}\gamma$. We find
\begin{align}
\delta_{1}\gamma
=\frac{\displaystyle \frac{1}{m_{1}^{2}}\frac{\delta \phi_{1}}{\phi_{1}}
-\frac{1}{m_{2}^{2}}\frac{\delta \phi_{2}}{\phi_{2}}}
{\displaystyle \frac{1}{m_{1}^{2}}\frac{\partial}{\partial\gamma }\ln\phi_{1,f}
-\frac{1}{m_{2}^{2}}\frac{\partial}{\partial\gamma }\ln\phi_{2,f}}
=\frac{\displaystyle \frac{m_{2}^{2}}{\phi_{1}}\delta \phi_{1}
-\frac{m_{1}^{2}}{\phi_{2}}\delta \phi_{2}}
{\displaystyle -m_{2}^{2}\frac{Y}{X}+m_{1}^{2}\frac{W}{Z}}\,.
\label{1gamma}
\end{align}
Here, for notational simplicity, we have introduced $X$, $Y$, $Z$ and $W$, 
which are defined by
\begin{eqnarray}
X&=&\frac{g_1g_2}{g\,\sigma}\,\phi_{1,f}
=\frac{1}{g}(g_{2}\cos\alpha \cos\gamma -g_{1}\sin\alpha \sin\gamma)\,, 
\cr
\cr
Y&=&-\frac{g_1g_2}{g\,\sigma}\,\frac{\partial}{\partial\gamma}\phi_{1,f}
=\frac{1}{g}(g_{2}\cos\alpha \sin\gamma +g_{1}\sin\alpha \cos\gamma)\,,
\cr
\cr
Z&=&\frac{g_1g_2}{g\,\sigma}\,\phi_{2,f}
=\frac{1}{g}(g_{2}\sin\alpha \cos\gamma +g_{1}\cos\alpha \sin\gamma)\,,
\cr
\cr
W&=&-\frac{g_1g_2}{g\,\sigma}\,\frac{\partial}{\partial\gamma}\phi_{2,f}
=\frac{1}{g}(g_{2}\sin\alpha \sin\gamma -g_{1}\cos\alpha \cos\gamma)\,,
\end{eqnarray}
where $g=\sqrt{g_1^2+g_2^2}$. The factor $1/g$ in front of each of
these quantities has been inserted for later convenience.

Then, collecting the second-order terms in Eq.~(\ref{expand}), we find
\begin{align}
\delta_{2}\gamma
&=\frac{1}{2}
\frac{\displaystyle-\frac{\delta\phi_{1}^{2}}{m_{1}^{2}\phi_{1}^{2}}
+\frac{\delta\phi_{2}^{2}}{m_{2}^{2}\phi_{2}^{2}}
-\left(
\frac{1}{m_{1}^{2}}\frac{\partial^{2}}{\partial\gamma^{2}}\ln\phi_{1,f}
-\frac{1}{m_{2}^{2}}\frac{\partial ^{2}}{\partial\gamma^{2}}\ln\phi_{2,f}
\right)(\delta_{1} \gamma )^{2}}
{\displaystyle\left(\frac{1}{m_{1}^{2}}
\frac{\partial }{\partial \gamma }\ln\phi_{1,f}
-\frac{1}{m_{2}^{2}}\frac{\partial }{\partial \gamma }\ln\phi_{2,f}\right)}
\notag\\
\notag\\
&=\frac{1}{2}
\frac{\displaystyle \left( \frac{m_{2}}{\phi_{1}}\delta\phi_{1}\right)^{2}
-\left( \frac{m_{1}}{\phi_{2}}\delta\phi_{2}\right)^{2}
+\left(-m_{2}^{2}\frac{X^2+Y^2}{X^2}
+m_{1}^{2}\frac{Z^2+W^2}{Z^2}\right)(\delta_{1}\gamma)^{2}}
{\displaystyle m_{2}^{2}\frac{Y}{X}-m_{1}^{2}\frac{W}{Z}}\,,
\label{2gamma}
\end{align}
where we note that
\begin{eqnarray}
X^2+Y^2=\frac{g_2^2}{g^2}\cos^2\alpha+\frac{g_1^2}{g^2}\sin^2\alpha\,,
\quad
Z^2+W^2=\frac{g_2^2}{g^2}\sin^2\alpha+\frac{g_1^2}{g^2}\cos^2\alpha\,,
\end{eqnarray}

Now, we compute $\delta N$. Although it is straightforward to
expand Eq.~(\ref{efolds}) to the second order in the field fluctuations,
the calculation is much simpler if we take the perturbation of either
of the solutions $\phi_1$ or $\phi_2$ of the slow roll equations of
motion~(\ref{slowrolleq}). For example, the solution for $\phi_1$ is
expressed as
\begin{eqnarray}
\phi_1=\phi_{1,f}e^{m_1^2N}
\quad
\leftrightarrow
\quad N=\frac{1}{m_1^2}\left(\ln\phi_1-\ln\phi_{1,f}\right)\,.
\end{eqnarray}
The perturbation of the second equation gives
\begin{eqnarray}
\delta N=\frac{1}{m_1^2}
\left[\frac{\delta\phi_1}{\phi_1}
-\frac{\partial}{\partial\gamma}\ln\phi_{1,f}\,\delta_1\gamma
-\frac{1}{2}\left(\frac{\delta\phi_1}{\phi_1}\right)^2
-\frac{1}{2}\frac{\partial^2}{\partial\gamma^2}\ln\phi_{1,f}\,(\delta_1\gamma)^2
-\frac{\partial}{\partial\gamma}\ln\phi_{1,f}\,\delta_2\gamma
\right]\,.
\label{Ndelta}
\end{eqnarray}
Inserting Eqs.~(\ref{1gamma}) and (\ref{2gamma}) into Eq.~(\ref{Ndelta}),
we obtain
\begin{gather}
\delta N
=\frac{\displaystyle
-\frac{W}{Z}\frac{\delta\phi_{1}}{\phi_{1}}
+\frac{Y}{X}\frac{\delta\phi_{2}}{\phi_{2}}}
{\displaystyle m_{2}^{2}\frac{Y}{X}-m_{1}^{2}\frac{W}{Z}}
+\frac{1}{2}\frac{\displaystyle
 \frac{W}{Z}\left(\frac{\delta\phi_{1}}{\phi_{1}}\right)^{2}
-\frac{Y}{X}\left(\frac{\delta\phi_{2}}{\phi_{2}}\right)^{2}}
{\displaystyle
m_{2}^{2}\frac{Y}{X}-m_{1}^{2}\frac{W}{Z}}
-\frac{1}{2}\frac{\displaystyle
\left( 1-\frac{YW}{XZ}\right)\left(\frac{W}{Z}-\frac{Y}{X}\right)
\left(\frac{m_{2}^{2}}{\phi_{1}}\delta \phi_{1}
-\frac{m_{1}^{2}}{\phi_{2}}\delta \phi_{2}\right)^{2}}
{\displaystyle
\left(m_{2}^{2}\frac{Y}{X}-m_{1}^{2}\frac{W}{Z}\right)^{3}}\,.
\end{gather}

Finally, we mention that we can divide $\delta N$ into two
contributions: one from during inflation up to a surface of
constant potential energy, $\delta N_*$, and 
the contribution from the end of inflation, $\delta N_e$. 
In the case of the exponential potential model
considered in~\cite{Sasaki:2008uc}, there was no non-Gaussianity
in $\delta N_*$ to the lowest order in the slow-roll parameters.
In contrast, there exists non-Gaussianity in $\delta N_*$
in the present model. Nevertheless, it can be easily shown that
it is of the order of the slow-roll parameters, and hence is negligibly
small.

\section{Linear Exponential Potential Model}
\label{app:linear}

In this appendix, we consider the case of a linear
exponential potential,
\begin{gather}
V=V_{0}\exp (m_{1}\phi_{1}+m_{2}\phi_{2})\,,
\end{gather}
with the condition for the end of inflation given by
\begin{align}
\sigma ^{2}
=g_{1}^{2}(\phi_{1}\cos\alpha+\phi_{2}\sin\alpha)^{2}
+g_{2}^{2}(-\phi_{1}\sin\alpha+\phi_{2}\cos\alpha)^{2}\,.
\label{gencond}
\end{align}
This model was discussed in~\cite{Sasaki:2008uc}.
However, it was assumed that $\alpha=0$. 
Here, for the sake of completeness, we consider the 
general condition adopted in the main text.

As in \S 2, we parametrize the scalar 
field at the end of inflation as
\begin{align}
\frac{\sigma}{g_{1}}\cos\gamma=\phi_{1,f}\cos\alpha+\phi_{2,f}\sin\alpha\,,
\quad
\frac{\sigma}{g_{2}}\sin\gamma=-\phi_{1,f}\sin\alpha+\phi_{2,f}\cos\alpha\,,
\end{align}
or, conversely,
\begin{gather}
\phi_{1,f}=\frac{\sigma}{g_{1}g_{2}}
(g_{2}\cos\alpha \cos\gamma-g_{1}\sin\alpha \sin\gamma)\,,
\quad
\phi_{2,f}=\frac{\sigma}{g_{1}g_{2}}
(g_{2}\sin\alpha \cos\gamma+g_{1}\cos\alpha \sin\gamma)\,.
\end{gather}
Also, as before, we introduce $g=\sqrt{g_1^2+g_2^2}$,
and $X$, $Y$, $Z$ and $W$ as
\begin{eqnarray}
X&=&\frac{g_1g_2}{g\,\sigma}\,\phi_{1,f}
=\frac{1}{g}(g_{2}\cos\alpha \cos\gamma -g_{1}\sin\alpha \sin\gamma)\,, 
\cr
\cr
Y&=&-\frac{g_1g_2}{g\,\sigma}\,\frac{\partial}{\partial\gamma}\phi_{1,f}
=\frac{1}{g}(g_{2}\cos\alpha \sin\gamma +g_{1}\sin\alpha \cos\gamma)\,,
\cr
\cr
Z&=&\frac{g_1g_2}{g\,\sigma}\,\phi_{2,f}
=\frac{1}{g}(g_{2}\sin\alpha \cos\gamma +g_{1}\cos\alpha \sin\gamma)\,,
\cr
\cr
W&=&-\frac{g_1g_2}{g\,\sigma}\,\frac{\partial}{\partial\gamma}\phi_{2,f}
=\frac{1}{g}(g_{2}\sin\alpha \sin\gamma -g_{1}\cos\alpha \cos\gamma)\,.
\end{eqnarray}

Let us calculate the curvature perturbation for this model.
To begin with, we evaluate the perturbation in $\gamma$ to the second order
to obtain
\begin{gather}
\delta_{1}\gamma 
=-\frac{g_{1}g_{2}}{g\,\sigma}
\frac{m_{2}\delta\phi_{1}-m_{1}\delta\phi_{2}}{m_{2}Y-m_{1}W}\,,
\quad
\delta_{2}\gamma=-\frac{(\delta_{1}\gamma)^{2}}{2}
\frac{m_{2}X-m_{1}Z}{m_{2}Y-m_{1}W}\,.
\end{gather}
On the basis of these equations, $\delta N$ is evaluated to the second order as
\begin{eqnarray}
\delta N=\frac{-W\delta\phi_{1}+Y\delta\phi_{2}}{m_{2}Y-m_{1}W}
+\frac{(g_{1}g_{2})^{2}}{2g^3\sigma}
\frac{(m_{2}\delta\phi_{1}-m_{1}\delta\phi_{2})^{2}}{(m_{2}Y-m_{1}W)^{3}}\,.
\end{eqnarray}

Now we can evaluate the quantities of interest. 
As before, for convenience, we introduce angle $\beta$ as
\begin{eqnarray}
g_1=g\cos\beta\,,\quad g_2=g\sin\beta\,.
\end{eqnarray}
Then the curvature perturbation spectrum is
\begin{gather}
{\cal P}_{S}=\frac{Y^2+W^2}{(m_{2}Y-m_1W)^2}
\left(\frac{H}{2\pi}\right)^2_{t_k}\,.
\end{gather}
The spectral index is
\begin{gather}
n_{S}=1-(m_{1}^{2}+m_{2}^{2})\,.
\end{gather}
The tensor-to-scalar ratio is
\begin{gather}
r=8\frac{(m_{2}Y-m_{1}W)^{2}}{Y^2+W^2}\,.
\end{gather}
Finally, the non-Gaussianity is
\begin{gather}
f_{NL}^{\rm local}
=\frac{5g}{6\sigma}
\frac{\cos^2\beta\sin^2\beta\,(m_{1}Y+m_{2}W)^{2}}
{(Y^2+W^2)^{2}(m_{2}Y-m_{1}W)}\,,
\end{gather}
where we note that
\begin{eqnarray}
Y^2+W^2=\cos^2\beta\cos^{2}\gamma+\sin^2\beta\sin^{2}\gamma\,.
\end{eqnarray}
Here, it is worthwhile to mention that the spectral index depends
only on $m_1$ and $m_2$.

To enable a direct comparison with the model discussed in the main text,
let us consider the case of $W=0$ for the present model as well.
In this case, we have
\begin{eqnarray}
{\cal P}_{S}&=&\frac{1}{m_{2}^2}\left(\frac{H}{2\pi}\right)^2_{t_k}\,,
\cr
\cr
r&=&8\,m_2^2\,,
\cr
\cr
f_{NL}^{\rm local}
&=&\frac{5g}{6\sigma}\frac{\cos^2\beta\sin^2\beta}
{(\cos^2\beta\cos^2\gamma+\sin^2\beta\sin^2\gamma)^{3/2}}
\frac{m_{1}^{2}}{m_{2}}\,.
\label{Wzerocase}
\end{eqnarray}
We see that a large mass ratio, $m_1\gg m_2$, is necessary 
in order to realize a large $f_{NL}^{\rm local}$.
However, because $r$ in the present case is determined only
by the smaller mass, $r=8\,m_2^2$, it is difficult to realize
both large $r$ and large $f_{NL}^{\rm local}$ .
This is in contrast to the case we discussed in
the main text, for which it was possible to make both
values large enough to be detectable in the very near future.


\begin{thebibliography}{18}


\bibitem{Salopek:1990jq}
  D.~S.~Salopek and J.~R.~Bond,
  Phys.\ Rev.\  D {\bf 42} (1990), 3936 .

\bibitem{Komatsu:2001rj}
  E.~Komatsu and D.~N.~Spergel,
  Phys.\ Rev.\ D {\bf 63} (2001), 063002; 
  astro-ph/0005036.

\bibitem{Babich:2004yc}
  D.~Babich and M.~Zaldarriaga,
  Phys.\ Rev.\  D {\bf 70} (2004), 083005; 
  astro-ph/0408455.


%

\bibitem{NGmodels1}
  F.~Bernardeau and J.~P.~Uzan,
  Phys.\ Rev.\  D {\bf 66} (2002), 103506; 
  hep-ph/0207295.
\\
  N.~Bartolo, S.~Matarrese and A.~Riotto,
  Phys.\ Rev.\ D {\bf 69} (2004), 043503; 
  hep-ph/0309033.
\\
  C.~Gordon and K.~A.~Malik,
  Phys.\ Rev.\ D {\bf 69} (2004), 063508; 
  astro-ph/0311102.
\\
  K.~Enqvist and S.~Nurmi,
  J. Cosmol. Astropart {\bf 0510} (2005), 013; 
  astro-ph/0508573.
\\
  D.~H.~Lyth,
  Nucl.\ Phys.\ Proc.\ Suppl.\  {\bf 148} (2005), 25.
\\
  K.~A.~Malik and D.~H.~Lyth,
  J. Cosmol. Astropart {\bf 0609} (2006), 008; 
  astro-ph/0604387.
\\
  M.~Sasaki, J.~Valiviita and D.~Wands,
  Phys.\ Rev.\ D {\bf 74} (2006), 103003; 
  astro-ph/0607627.
\\
  J.~Valiviita, M.~Sasaki and D.~Wands,
  astro-ph/0610001.
\\
  S.~Yokoyama, T.~Suyama and T.~Tanaka,
  J. Cosmol. Astropart {\bf 0707} (2007), 013; 
  arXiv:0705.3178.
\\
  S.~Yokoyama, T.~Suyama and T.~Tanaka,
  arXiv:0711.2920.
\\
  K.~Ichikawa, T.~Suyama, T.~Takahashi and M.~Yamaguchi,
  arXiv:0802.4138.
\\
  T.~Suyama and F.~Takahashi,
  arXiv:0804.0425.
\\
  T.~Matsuda,
  arXiv:0804.3268.
\\
  F.~Bernardeau and T.~Brunier,
  Phys.\ Rev.\  D {\bf 76} (2007), 043526; 
  arXiv:0705.2501.
\\
  D.~A.~Easson, R.~Gregory, D.~F.~Mota, G.~Tasinato and I.~Zavala,
  J. Cosmol. Astropart {\bf 0802} (2008), 010; 
  arXiv:0709.2666.

\bibitem{NGmodels2}
  M.~Zaldarriaga,
  Phys.\ Rev.\ D {\bf 69} (2004), 043508; 
  astro-ph/0306006.
\\
  G.~N.~Felder and L.~Kofman,
  hep-ph/0606256.
\\
 T.~Suyama and M.~Yamaguchi,
 Phys.\ Rev.\  D {\bf 77} (2008), 023505; 
 arXiv:0709.2545.
\\
  F.~Bernardeau, L.~Kofman and J.~P.~Uzan,
  Phys.\ Rev.\  D {\bf 70} (2004), 083004; 
  astro-ph/0403315.

\bibitem{Bernardeau:2002jf}
  F.~Bernardeau and J.~P.~Uzan,
  Phys.\ Rev.\  D {\bf 67} (2003), 121301; 
  astro-ph/0209330.

\bibitem{Dvali:2003em}
 G.~Dvali, A.~Gruzinov and M.~Zaldarriaga,
 Phys.\ Rev.\  D {\bf 69} (2004), 023505; 
 astro-ph/0303591.

\bibitem{Kofman:2003nx}
 L.~Kofman,
 astro-ph/0303614.

\bibitem{Lyth:2005qk}
  D.~H.~Lyth,
  J. Cosmol. Astropart {\bf 0511} (2005), 006; 
  astro-ph/0510443.


\bibitem{Alabidi}
  L.~Alabidi and D.~Lyth,
  J. Cosmol. Astropart {\bf 0608} (2006), 006; 
  astro-ph/0604569.
\\
  L.~Alabidi,
  J. Cosmol. Astropart {\bf 0610} (2006), 015; 
  astro-ph/0604611.

\bibitem{BarnabyCline}
  N.~Barnaby and J.~M.~Cline,
  Phys.\ Rev.\  D {\bf 75} (2007), 086004; 
  astro-ph/0611750.
\\
  N.~Barnaby and J.~M.~Cline,
  J. Cosmol. Astropart {\bf 0707} (2007), 017; 
  arXiv:0704.3426.
\\
  N.~Barnaby and J.~M.~Cline,
  J. Cosmol. Astropart {\bf 0806} (2008), 030; 
  arXiv:0802.3218.

\bibitem{Chambers:2007se}
  A.~Chambers and A.~Rajantie,
  Phys.\ Rev.\ Lett.\  {\bf 100} (2008), 041302; 
  arXiv:0710.4133.

\bibitem{Silverstein:2003hf}
  E.~Silverstein and D.~Tong,
  Phys.\ Rev.\  D {\bf 70} (2004), 103505; 
  hep-th/0310221.

\bibitem{DBIinflation}
  M.~Alishahiha, E.~Silverstein and D.~Tong,
  Phys.\ Rev.\  D {\bf 70} (2004), 123505; 
  hep-th/0404084.
\\
  D.~Seery and J.~E.~Lidsey,
  J. Cosmol. Astropart {\bf 0506} (2005), 003; 
  astro-ph/0503692.
\\
  X.~Chen, M.~X.~Huang, S.~Kachru and G.~Shiu,
  J. Cosmol. Astropart {\bf 0701} (2007), 002; 
  hep-th/0605045.
\\
  M.~X.~Huang, G.~Shiu and B.~Underwood,
  Phys.\ Rev.\  D {\bf 77} (2008), 023511; 
  arXiv:0709.3299.
\\
  D.~Langlois, S.~Renaux-Petel, D.~A.~Steer and T.~Tanaka,
  arXiv:0804.3139.
  arXiv:0806.0336.
\\
  F.~Arroja, S.~Mizuno and K.~Koyama,
  arXiv:0806.0619.

\bibitem{Sasaki:2008uc}
  M.~Sasaki,
  Prog.\ Theor.\ Phys.\  {\bf 120} (2008), 159; 
  arXiv:0805.0974 .

\bibitem{Sasaki:1995aw}
  M.~Sasaki and E.~D.~Stewart,
  Prog.\ Theor.\ Phys.\  {\bf 95} (1996), 71; 
  astro-ph/9507001.

\bibitem{Sasaki:1998ug}
  M.~Sasaki and T.~Tanaka,
  Prog.\ Theor.\ Phys.\  {\bf 99} (1998), 763; 
  gr-qc/9801017.

\bibitem{Wands:2000dp}
  D.~Wands, K.~A.~Malik, D.~H.~Lyth and A.~R.~Liddle,
  Phys.\ Rev.\ D {\bf 62} (2000), 043527; 
  astro-ph/0003278.

\bibitem{Lyth:2004gb}
  D.~H.~Lyth, K.~A.~Malik and M.~Sasaki,
  J. Cosmol. Astropart {\bf 0505} (2005), 004; 
  astro-ph/0411220.

\bibitem{Lyth:2005fi}
  D.~H.~Lyth and Y.~Rodriguez,
  Phys.\ Rev.\ Lett.\  {\bf 95} (2005), 121302; 
  astro-ph/0504045.

\bibitem{Cogollo:2008bi}
  H.~R.~S.~Cogollo, Y.~Rodriguez and C.~A.~Valenzuela-Toledo,
  J. Cosmol. Astropart {\bf 0808} (2008), 029; 
  arXiv:0806.1546.

\bibitem{Sasaki:2007ay}
  M.~Sasaki,
  Class.\ Quantum.\ Grav.\  {\bf 24} (2007), 2433; 
  astro-ph/0702182.

\bibitem{Starobinsky:1986fxa}
  A.~A.~Starobinsky,
  JETP Lett.\  {\bf 42} (1985) 152
  [Pisma Zh.\ Eksp.\ Teor.\ Fiz.\  {\bf 42} (1985) 124].

\bibitem{Liddle:2006ev}
 A.~R.~Liddle, D.~Parkinson, S.~M.~Leach and P.~Mukherjee,
 Phys.\ Rev.\  D {\bf 74} (2006), 083512; 
 astro-ph/0607275.


\bibitem{Dunkley:2008ie}
  J.~Dunkley {\it et al.}  (WMAP Collaboration),
  arXiv:0803.0586.

\bibitem{Komatsu:2008hk}
  E.~Komatsu {\it et al.}  (WMAP Collaboration),
  arXiv:0803.0547.

\end{thebibliography}
\end{document}